\documentclass[twocolumn]{article}
\usepackage{tabularray}
\usepackage{ninecolors}
\usepackage{setspace}
\usepackage{hyperref}
\usepackage{float}
\usepackage{parskip}
\usepackage[margin=1.5cm]{geometry}
\usepackage{graphicx}
\usepackage{rotating}
\UseTblrLibrary{booktabs}
\usepackage{pdflscape}
\usepackage{soul}
\usepackage{verbatim}
\usepackage[switch]{lineno}
\usepackage{cuted}
\usepackage[superscript]{cite}

\title{Political polarization and mental wellbeing: asymmetric evidence for bidirectionality}
\author{
Afrooz Mahir\textsuperscript{1}*,
Shivam Sharma\textsuperscript{1},
Juhi Kulshrestha\textsuperscript{1},
Mikko Kivelä\textsuperscript{1},
Talayeh Aledavood\textsuperscript{1}\\
\textsuperscript{1}Department of Computer Science, Aalto University, Espoo, Finland
}
\date{}

\begin{document}

\twocolumn[
\maketitle
\begin{abstract}

A growing body of evidence suggests that political polarization and mental health and wellbeing influence each other. Yet the two directions, how polarization shapes mental wellbeing and how mental wellbeing shapes polarization, remain siloed across disciplines. We review both literatures to evaluate the strength and alignment of evidence for bidirectional effects. We find that the literature varies widely in constructs, measures, and levels of analysis, and that the evidence for the two pathways is asymmetric: links from polarization to mental wellbeing are more direct, grounded in well-developed theoretical frameworks and social-relational mechanisms. In contrast, the reverse pathway is substantially less direct, with studies often examining cognitive and socio-emotional factors, or political outcomes adjacent to polarization rather than polarization itself. We identify clear gaps in the literature highlighting the need for future studies with conceptual clarity, unified frameworks, validated measures, and multilevel designs examining both pathways within the same empirical and theoretical setting.

\end{abstract}
\vspace{25pt}
]

Among the social and political challenges faced by contemporary societies, increasing political polarization and declining mental wellbeing are particularly consequential. Polarization, expressed through increasingly opposed political views and growing hostility between groups, can erode social and institutional trust, hinder political cooperation, and deepen social division \cite{tucker_social_2018, mccoy_polarization_2018, iyengar_origins_2019, finkel_political_2020, mccoy_toward_2019}. At the same time, mental health problems such as depression and anxiety impair individual wellbeing and social functioning \cite{noauthor_mental_nodate}. These phenomena can also be connected: political polarization may lead to poorer mental wellbeing among individuals experiencing it, while psychological characteristics and mental states may influence how individuals interpret and respond to political differences, in turn increasing polarization. 
Understanding these links is therefore important for individual mental wellbeing and for social cohesion and democratic functioning.

A substantial body of literature reports how polarization can lead to worsening mental wellbeing. Some studies rely on population-level associations: individuals who perceive greater partisan division report more days of poor mental and physical health \cite{fraser_harmful_2022}, a higher likelihood of developing anxiety, depression, and sleep disorders \cite{nayak_is_2021}, and lower social trust and willingness to cooperate \cite{lee_social_2022}. Others explore individual-level mechanisms, such as sustained political division shaping wellbeing through perceived threat, social exclusion, identity conflict, and weakened social support, all of which are closely linked to mental health and wellbeing
(hereafter, mental wellbeing)~\cite{jost_cognitivemotivational_2022, berkman_social_2000, baumeister_need_nodate}.Furthermore, a growing body of evidence suggests that political polarization may function as a chronic psychosocial stressor \cite{neely_anxious_2026, fraser_harmful_2022, nayak_is_2021}.

A separate body of literature points to a possible reverse pathway. For example, emotional states and psychological traits, such as cognitive rigidity and narcissistic rivalry, along with negative interpersonal experiences, may shape individuals' engagement with and contribution to political polarization \cite{zmigrod_role_2020, tilley_narcissism_2025, kam_hurt_2025}. Mental wellbeing trajectories and associated cognitive processes may also shape political preferences and orientations, with internalizing problems such as depression and anxiety symptoms linked to distinct political attitudes and patterns of political thinking \cite{fluit_mental_2026, edinger_cognitive_2025}. However, this literature remains smaller and conceptually broader than the literature on the first pathway.

Together, the findings from these two literatures raise the possibility of a bidirectional relationship, in which political polarization shapes mental wellbeing while psychological traits and states, in turn, contribute to polarization. Understanding this bidirectionality requires clarifying the extent to which existing findings reflect mere associations rather than causal effects, how strong the causal links are in each direction, and whether the two directions can be distinguished from one another. Most critically, the mechanisms underlying both directions have yet to be examined directly.

Clarifying which of these patterns are most prominent has important implications for both research and interventions. If polarization predominantly drives mental wellbeing outcomes, reducing polarization should lead to downstream psychological benefits; if the reverse is true, addressing mental wellbeing issues may reduce political extremism. If the relationship is genuinely bidirectional, cumulative feedback loops could produce escalating cycles in which polarization and reduced mental wellbeing reinforce one another, making early intervention critical before such dynamics consolidate. Alternatively, observed associations may reflect shared contextual drivers: for instance, social isolation can affect both polarization and mental wellbeing while being influenced by broader social changes or external shocks such as the COVID-19 pandemic \cite{me_hurt_nodate, van2024political}, and in this case targeting either domain alone would be insufficient without addressing the common underlying causes.

Evaluating this bidirectional relationship is complicated by a substantial conceptual mismatch between the literatures describing the two pathways. Studies examining polarization as a predictor typically operationalize it as ideological distance, partisan affect, perceived societal division, or network segregation \cite{iyengar_affect_2012, yarchi_political_2021, bakshy_exposure_2015}, and assess outcomes ranging from validated measures of depression and anxiety \cite{spitzer_brief_2006} to subjective wellbeing, social connectedness, and computational linguistic proxies. By contrast, research relevant to the reverse pathway more often examines cognitive rigidity, ideological extremity, distrust, radicalization, or related political attitudes, rather than directly measuring both mental wellbeing and political polarization \cite{kubin_role_2021}. The two pathways therefore differ not only in the amount and type of available evidence, but also in the constructs, measures, and levels of analysis on which that evidence rests. Consequently, findings that appear consistent with bidirectionality are often not directly comparable.

\begin{figure}[h]
    \centering
    \includegraphics[width=1\linewidth]{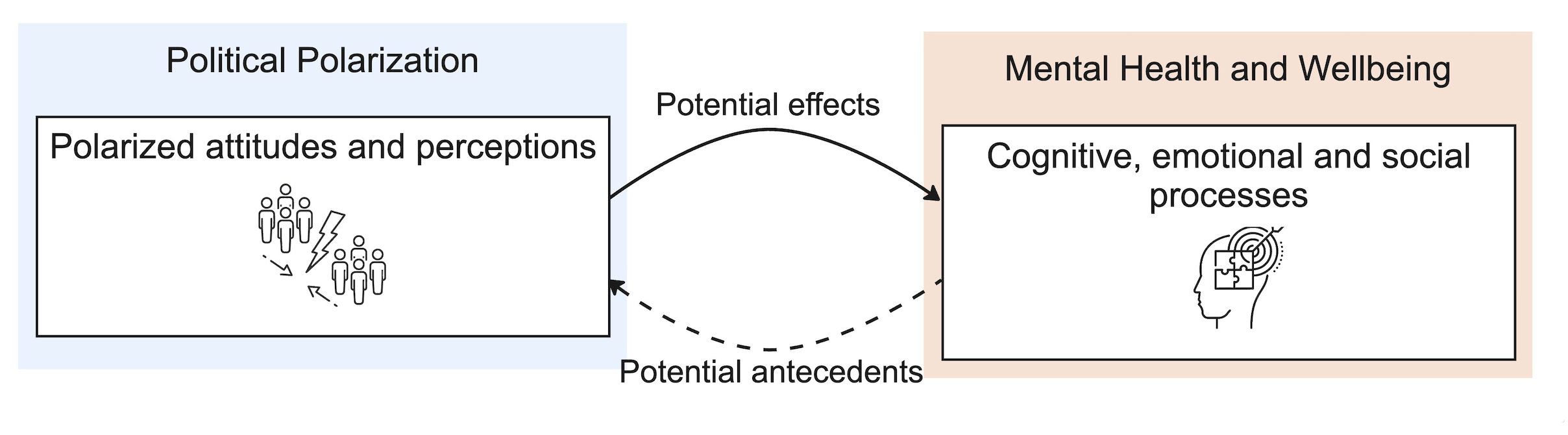}
    \caption{\textbf{Conceptual model of the potential bidirectional relationship between political polarization and mental wellbeing.} Solid lines indicate more direct evidence, whereas dashed lines indicate more indirect or heterogeneous evidence. The two pathways differ in evidentiary strength: evidence is more direct for associations from polarization to mental wellbeing outcomes, whereas the reverse pathway more often relies on broader psychological antecedents and polarization-related outcomes.}
    \label{fig:bidirectional}
\end{figure}

In this review, we evaluate evidence for a bidirectional relationship between political polarization and mental wellbeing (Figure \ref{fig:bidirectional}). We first examine forms of political polarization, their measurement,
and their associations with mental wellbeing outcomes, together with the social and psychological processes through which these associations may arise. We then assess the reverse pathway, distinguishing evidence that mental wellbeing states predict polarization from the broader literature linking cognitive, affective, and personality traits to polarization. Across both pathways, we distinguish direct evidence from evidence based on mediating processes, psychological antecedents, contextual indicators, and computational proxies. Distinguishing these forms of evidence allows us to assess where evidence for bidirectionality is strongest, where it remains indirect or conceptually mismatched, and which conceptual and methodological gaps currently prevent more robust testing of reciprocal relationships.

\section*{Methodology}
This study was conducted as a narrative review using an integrative approach. A narrative review was considered appropriate because research on the relationship between political polarization and mental wellbeing spans multiple disciplines and uses diverse conceptualizations and measurement approaches.

A structured, non-systematic literature search was conducted primarily using Google Scholar and Scopus, with searches conducted in October 2025 and updated in August 2026. Searches included combinations of terms related to political polarization (e.g., political polarization, partisan conflict, affective polarization, ideological polarization, perceived polarization) and mental wellbeing (e.g., mental health, psychological wellbeing, anxiety, depression, stress, cognition, and wellbeing). Searches were limited to studies published in English or studies for which an English translation was available. Additional relevant studies were identified through reference list screening and citation tracking. Studies were included if they examined associations between political polarization and mental wellbeing, investigated psychological mechanisms relevant to either direction of the relationship, or provided theoretical insights relevant to understanding these links. Studies focusing on non-political forms of polarization or lacking relevance to psychological outcomes were excluded. A total of 34 studies were included in the final narrative synthesis.

Included studies were analyzed using thematic synthesis, with evidence organized according to the direction of the association (\textit{political polarization → mental wellbeing} and \textit{mental wellbeing → political polarization}), the type or indicator of polarization examined, and the role of the psychological variable within the proposed relationship. Psychological variables were distinguished as direct wellbeing outcomes, social-psychological mediators, cognitive and personality factors, or computational proxies, allowing evidence with different inferential status to be compared without treating these constructs as equivalent measures of mental wellbeing. This approach allowed integration of findings across psychology, political science, and related disciplines while accounting for differences in conceptual definitions and measurement strategies.

\section*{Conceptualization and measurement of political polarization} 

\subsection*{Forms of political polarization}
Political polarization can take multiple forms and can be conceptualized at individual, relational, and societal levels. Here, we focus on the forms most consistently operationalized in political polarization research and most relevant to the literature on mental wellbeing. Other distinctions, including positional, epistemic, geographic, moral, and intransigent polarization, are considered where relevant but are not treated as separate analytical categories. Table \ref{tab:framework_polarization} summarizes the principal forms of political polarization represented in the reviewed literature and their hypothesized or observed relevance to mental wellbeing.

Political polarization manifests in several conceptually distinct forms, summarized with full definitions in Table \ref{tab:framework_polarization}. \textit{Ideological} and \textit{affective polarization} are conceptually distinct \cite{wilson_polarization_2020, iyengar_origins_2019, druckman_what_2019, kubin_role_2021}: the former reflects divergence in policy positions and beliefs, while the latter centers on hostility toward out-groups and favouritism toward in-groups, defined primarily by negative emotions, distrust, and social distancing \cite{mcmurtrie_development_2024}. \textit{Perceived polarization}, individuals' perception of political distance between groups, can reflect either accurate perceptions or exaggerated ones (\textit{false polarization}) \cite{yang_why_2016, wilson_polarization_2020}, and such misperception intensifies resentment even when objective divisions are moderate. Online, partisan echo chambers and homophilic network ties can further amplify perceptions of division \cite{kubin_role_2021, frimer_incivility_2023, fernbach_false_2022, moore-berg_exaggerated_2020}.
Offline measures of polarization have traditionally relied on survey-based methods, which capture both ideological and affective dimensions. Ideological polarization is commonly measured using ideology scales or policy-specific items, while evaluative measures such as warmth, favourability, or attitudes toward political leaders are also used \cite{kubin_role_2021}. Affective polarization is commonly assessed using feeling thermometers, trait evaluations, and social distance measures, which capture emotional warmth, negative evaluations, stereotyping, and willingness to interact with political out-groups \cite{iyengar_origins_2019, druckman_what_2019, lelkes_mass_2016, mcmurtrie_development_2024}. These measures vary in what they capture, complicating cross-study comparisons \cite{kubin_role_2021}, although newer multidimensional scales incorporate related dimensions such as social distance, aversion, incivility, othering, and moralization \cite{mcmurtrie_development_2024,campos_new_2025}. 
In digital contexts, additional forms have been identified: \textit{interactional} or \textit{network polarization}, the tendency for online networks to divide into ideologically homophilic clusters that reduce opportunities for inter-group dialogue \cite{yarchi_political_2021, eck_opposing_2024, salloum_separating_2022, falkenberg_patterns_2024}; \textit{interpretive polarization}, divergent and often delegitimizing interpretations of political issues \cite{kligler-vilenchik_interpretative_2020}; toxic or hostile online expression \cite{marchal_be_2022, falkenberg_patterns_2024}; and, at the extreme, \textit{destructive} or \textit{pernicious polarization}, in which opposing groups view one another as existential threats, undermining democratic norms, institutional trust, and social cohesion \cite{mccoy_toward_2019,somer_pernicious_2023,mccoy_polarization_2018,esau_destructive_2025}. Evidence for fully self-reinforcing ``filter bubbles'' or ``echo chambers,'' however, remains limited; users show a preference for like-minded content but are still exposed to opposing views to some degree \cite{ross_arguedas_echo_2022,tornberg_how_2022, eck_opposing_2024}. Overall, political polarization encompasses ideological, affective, and identity-based dimensions, reflecting both attitudes and emotional responses toward political groups, and is measured using diverse survey-based, behavioral, and digital approaches that capture perceived divisions, inter-group attitudes, and patterns of interaction; this diversity also complicates comparisons across studies.

\begin{table*}[!h]
\small
\centering
\caption{Types of political polarization and their relevance to mental wellbeing}
\label{tab:framework_polarization}
\begin{tblr}[
]{
  colspec  = {Q[l,0.18\textwidth] Q[l,0.32\textwidth] Q[l,0.42\textwidth]},
  width    = \textwidth,
  rowhead  = 1,
  hlines   = {black},
  row{1}   = {font=\bfseries}
}
 
\textbf{Type} & \textbf{Definition} & \textbf{Relevance to Mental Wellbeing} \\
 
Ideological/issue-based polarization
& Divergence in broad political beliefs, values, or policy positions across groups, where opinions become increasingly separated and concentrated toward opposing ideological positions \cite{dimaggio_have_1996, dalton_modeling_2021}
& Driven by cognitive–motivational mechanisms, including ego and group justification processes \cite{jost_cognitivemotivational_2022}. Belief reinforcement within ideologically aligned networks may further maintain polarized views \cite{baumgaertner_preference_2022}, while cognitive distortion patterns, also observed in conditions such as depression and anxiety, have been identified in polarized political discourse \cite{edinger_cognitive_2025}. \\ 
 
Affective polarization
& In-group favouritism and out-group hostility based on partisan identity,
  characterised by antipathy, distrust, avoidance, and social distancing,
  often independent of ideological differences
  \cite{iyengar_affect_2012, iyengar_origins_2019, druckman_what_2019}
& Shaped by political emotions that affect inter-group relations and social inclusion \cite{von_scheve_putting_2024}, and is linked to reduced social support, elevated stress, and poorer health outcomes \cite{mcmurtrie_political_2025, bakker_putting_2024}. \\
 
False / Perceived polarization
& An individual's subjective overestimation of societal division; people
  systematically perceive partisan divides as larger than they objectively are
  \cite{fernbach_false_2022, blatz_false_2018}
& Shaped and intensified by underlying cognitive and affective tendencies, including categorical thinking, information simplification, and heightened emotional responses \cite{fernbach_false_2022}. It may contribute to psychological distress even when objective levels of political division are unchanged, and has been linked to an elevated likelihood of anxiety, depression, and sleep disorders \cite{nayak_is_2021, fraser_harmful_2022, panagopoulos_bridging_2022}. 
  
  \\

Interactional / network polarization
& Reduction in cross-cutting social ties and possible formation of partisan echo chambers
  in social and online networks; homophily-driven information exposure and
  partisan social sorting
  \cite{yarchi_political_2021, interian_network_2023, mepham_network_nodate, bakshy_exposure_2015, barbera_birds_2015} 
& Theorized to weaken social connectedness, reduce perceived belonging, and diminish social support, all of which are important determinants of psychological wellbeing \cite{berkman_social_2000, baumeister_need_nodate}.

\\

Interpretive polarization
& Divergent framing and meaning-making across partisan lines; opposing groups
  assign different interpretations to shared events or information
  \cite{kligler-vilenchik_interpretative_2020}
& May contribute to cognitive rigidity, biased information processing, and heightened uncertainty through divergent interpretations of shared political realities \cite{jost_cognitivemotivational_2022, salvi_does_2023}. \\

Destructive / pernicious polarization
& Extreme ``Us vs.\ Them'' divisions marked by mutual distrust and perceived threat that undermine democratic norms and social cohesion  \cite{mccoy_polarization_2018, mccoy_toward_2019}; captures the escalation of these dynamics across political conflicts \cite{esau_destructive_2025} & May lead to political despair, erosion of institutional trust, and support for political violence; hostile world perceptions linked to poorer mental health outcomes
  \cite{piazza_political_2023, kalmoe_threats_2024, pfundmair_how_2024} \\

\end{tblr}
\end{table*}

Digital polarization is also studied computationally, through content-based methods that analyze toxicity, sentiment, and partisan framing in political language \cite{falkenberg_patterns_2024,kubin_role_2021, karjus_evolving_2024}, network-based methods that map homophily and community structure in online interactions \cite{salloum_separating_2022,interian_network_2023,phillips_high-dimensional_2023}, and, increasingly, hybrid approaches combining both to capture the temporal and structural dynamics of online polarization, such as the Social Political Polarization analysis by INformation theory (SPIN) framework \cite{munoz_quantifying_2024,hohmann_quantifying_2023,matakos_measuring_2017}.

\subsection*{Proxy and indirect measures at the polarization–mental wellbeing intersection}
Across the intersection literature, political polarization is often not explicitly defined in terms of established subtypes such as affective, ideological, or perceived polarization. Instead, it is frequently operationalized indirectly through contextual indicators such as election periods, political crises, or individuals’ self-reported political orientation~\cite{relihan_election_2025, avidor_political_2025, zhao_2024_2025, pitcho-prelorentzos_factors_2018, gul_tekdal_life_2026, eshel_social_2026}. This approach assumes that heightened political conflict increases inter-group distance, intolerance, and emotionally salient responses to political events, which may affect mental health outcomes \cite{avidor_political_2025}. However, such studies capture exposure to politicized contexts rather than specific dimensions of polarization.

A similar issue appears in studies examining the reverse direction (\textit{mental wellbeing $\rightarrow$ polarization}), where polarization is not directly measured but approximated through related constructs such as ideological extremity, cognitive rigidity, polarized trust, conspiratorial beliefs, and socio-cognitive traits. While not equivalent to polarization, these constructs can be interpreted as proxies for cognitive, affective, and social mechanisms underlying polarization.

Several studies illustrate how psychological and cognitive constructs can capture aspects of political polarization without directly measuring polarization itself. For instance, Zmigrod \cite{zmigrod_role_2020} examines ideological extremity at the individual level, capturing the degree to which individuals hold extreme political beliefs rather than the distance between political groups. Pfundmair \cite{pfundmair_how_2024} links social exclusion to escalation processes involving perceived group threat, desensitization, and dehumanization within a broader radicalization framework, while Acar-Burkay et al. \cite{acar-burkay_trusting_2014} describe polarized trust judgments driven by cognitive closure and asymmetric trust between social groups. Similarly, Salvi et al. \cite{salvi_does_2023} conceptualize socio-cognitive polarization as a combination of ideological rigidity, intolerance of ambiguity, and xenophobia, spanning ideological and affective components. Baum et al. \cite{baum_political_2024} further associate conspiratorial beliefs and poor mental health with support for political violence, suggesting indirect links to affective or destructive polarization. 

Overall, polarization is frequently operationalized indirectly across both directions of the relationship, either through contextual political exposure or through psychologically adjacent proxy constructs. This limits comparability across studies and reflects a broader lack of conceptual consistency in how polarization is defined and measured within the intersection of political polarization and mental wellbeing research. It also highlights a wider set of psychological antecedents that may
underpin polarization-related outcomes.

\section*{Operationalization of mental wellbeing in polarization research}

\begin{table*}[!h]
\small
\centering
\caption{Psychological variables and operationalizations used to study mental wellbeing in polarization research}
\label{tab:framework_mh}
\begin{tblr}[
]{
  colspec  = {Q[l,0.22\textwidth] Q[l,0.48\textwidth] Q[l,0.22\textwidth] Q[l,0.20\textwidth]},
  width    = \textwidth,
  rowhead  = 1,
  hlines   = {black},
  row{1}   = {font=\bfseries}
}
 
\textbf{Category} & \textbf{Examples} & \textbf{Measurement Validity} \\
 
Clinical symptoms 
& Depression, generalized anxiety, panic disorder, sleep disorders, phobia
& Strongest; clinically validated \\
 
General psychological wellbeing 
& Life satisfaction, subjective wellbeing, self-rated mental health, psychological distress
& Moderate; widely used \\
 
Social-psychological mediators
& Social trust, belonging, perceived threat, loneliness, social support, social connectedness
& Moderate; mediators, not direct mental wellbeing outcomes \\
 
Cognitive \& personality factors
& Cognitive rigidity, need for closure, narcissism, social exclusion, integrative complexity, hurt feelings
& Moderate; often validated psychometric instruments\\
 
Computational proxies
& Sentiment expression, emotional language, cognitive distortion language, clinical language in online posts
& Weakest; no clinical validation; cannot ascertain individual-level mental wellbeing inference \\
 
\end{tblr}
\end{table*}

Psychological variables are operationalized very differently across the literature, reflecting both its disciplinary breadth and the different roles that psychological constructs play within proposed polarization–mental wellbeing relationships. We distinguish five recurring categories: clinical symptoms, general psychological wellbeing, social-psychological mediators, cognitive and personality factors, and computational proxies (Table \ref{tab:framework_mh}). These categories should not be interpreted as equivalent measures of mental wellbeing. Rather, they range from direct indicators of mental health and wellbeing to potential mediators or antecedents and indirect signals of psychological states.

Clinical symptoms are assessed using validated instruments such as the PHQ-9 and CES-D for depression, the GAD-7 for generalized anxiety, PSS-10 for perceived stress, and structured clinical interviews. These instruments provide clinically meaningful assessments, with some also enabling comparison against established clinical thresholds \cite{zhao_2024_2025, pitcho-prelorentzos_factors_2018, baum_political_2024, avidor_political_2025, eshel_social_2026, fluit_mental_2026, walker_emotional_2026, mcmurtrie_political_2025}. 

General psychological wellbeing is the most common measurement approach in survey-based studies. Indicators such as life satisfaction \cite{gul_tekdal_life_2026}, positive and negative affect \cite{smith_pins_2024, versteegen_we_2024, relihan_election_2025}, self-reported poor health days \cite{fraser_harmful_2022, nayak_is_2021, makri_anxiety_2024, panagopoulos_bridging_2022, nelson_resentment_2022, smith_politics_nodate}, and single-item self-rated mental health \cite{shepherd2026effects} capture broad wellbeing but do not map onto clinical conditions or diagnostic categories. While these measures are widely used and practical for large-scale surveys, their lack of clinical validation limits the strength of conclusions that can be drawn about mental health in any diagnostic sense.

Social psychological mediators, including social trust, social connectedness, belonging, and perceived threat, appear prominently as outcome variables in several studies, reflecting the theorized \textit{Polarization $\rightarrow$ mental wellbeing} pathway \cite{lee_social_2022, piazza_political_2023, reynolds_college_2025, gulevich_how_2026}. They are typically measured using interviews, experiments, and survey scales, which vary in their degree of validation across studies.

Cognitive and personality factors, including cognitive rigidity \cite{zmigrod_role_2020, wu_framework_2022, salvi_does_2023}, preference for belief \cite{baumgaertner_preference_2022}, radicalism \cite{pfundmair_how_2024}, need for closure \cite{acar-burkay_trusting_2014}, narcissism \cite{tilley_narcissism_2025}, and integrative complexity and hurt feelings \cite{kam_hurt_2025}, represent a distinct category that appears primarily in studies examining the \textit{mental wellbeing $\rightarrow$ Polarization}. These constructs function simultaneously as indicators of psychological functioning and as antecedents of political behavior, and are typically assessed using validated psychometric instruments such as the Cognitive Flexibility Inventory (CFI) \cite{dennis_cognitive_2010}, the  Kruglanski Need for Closure Scale (NFC-42) \cite{kruglanski_need_2009}, and the Narcissistic Admiration and Rivalry Questionnaire (NARQ) \cite{noauthor_narcissistic_nodate}. Their prominence in the \textit{mental wellbeing $\rightarrow$ Polarization} direction suggests that individual-level psychological characteristics may be important drivers of polarization that have been underexplored in the \textit{Polarization$\rightarrow$mental wellbeing} direction.

Computational proxies, including sentiment scores, emotional expression patterns, and cognitive distortion language derived from social media text, are the dominant approach in studies examining polarization in online environments. Traditional tools 
include VADER for valence-based sentiment scoring \cite{hutto_vader_2014} and LIWC \cite{tausczik_psychological_2010} for psycholinguistic profiling of emotional and cognitive language patterns \cite{pendse_role_2025}. More 
recent research has employed custom NLP classifiers trained to detect cognitive distortion language, such as catastrophizing, overgeneralizing, and emotional reasoning, as indirect proxies for psychological distress \cite{edinger_cognitive_2025}. However, none of these tools has been clinically validated against diagnostic standards, and population-level language patterns cannot be reliably interpreted as individual mental health conditions. This limitation is particularly important in polarization research, where emotionally charged political language may resemble expressions of distress without indicating actual psychological difficulties.

This variation in measurement creates two main challenges. First, it reduces comparability across studies, as survey-based measures and computational indicators capture different constructs and operate at different levels of analysis. Second, it limits causal interpretation because most studies using computational proxies are cross-sectional and correlational, making it unclear whether distress-related language precedes or follows polarization. The limited longitudinal, clinically grounded evidence and uneven coverage of polarization-specific dynamics represent key gaps, motivating research that integrates validated psychological measures, explicit polarization measures, and computational approaches across both directions.

\section*{Associations between political polarization and mental wellbeing}

\subsection*{Evidence and conceptual mismatch}

Table \ref{tab:polarization_wellbeing} brings the two directions of evidence together. A central difficulty is that the two pathways are supported by different kinds of evidence. Studies of \textit{polarization $\rightarrow$ mental wellbeing} more often assess recognizable forms or contextual indicators of polarization alongside mental wellbeing outcomes, whereas studies relevant to \textit{mental wellbeing $\rightarrow$ Polarization} frequently examine psychological antecedents alongside political attitudes, ideological extremity, or other polarization-adjacent outcomes. Consequently, apparent evidence for bidirectionality often rests on constructs that are neither conceptually nor operationally equivalent across the two directions.

\small
\onecolumn
\begin{longtblr}[
    caption={Studies examining associations between political polarization and mental wellbeing, summarised by polarization measures or indicators, mental wellbeing measures, and reported mechanisms or proxies.},
    label={tab:polarization_wellbeing}
]{
    width=\textwidth,
    colspec={
        X[2,l]
        X[1.8,l]
        X[2,l]
        X[0.6,l]
    },
    rowhead=1,
    hlines,
    row{1}={font=\bfseries},
    cells={valign=t}
}

\textbf{Polarization measure or indicator} &
\textbf{Mental wellbeing measure} &
\textbf{Mechanisms and proxies} &
\textbf{References} \\

\SetCell[c=4]{c}{\textbf{Polarization $\rightarrow$ Mental wellbeing}} \\

Perceived political distance, perceived partisan polarization, and state or national polarization
&
Poor mental health days, anxiety, depression, and sleep disorders
&
Political distance and polarization as contextual stressors; social capital
&
\cite{fraser_harmful_2022,nayak_is_2021,panagopoulos_bridging_2022} \\

Perceived and subjective polarization
&
Social trust, social connectedness, belonging, and psychological distress
&
Social trust, belonging, and social connectedness as potential mediating mechanisms
&
\cite{lee_social_2022,reynolds_college_2025} \\

Affective polarization, including partisan hostility, social distance, and negative emotions toward political out-groups
&
Self-rated health, perceived stress, and social support
&
Political participation, social support, and stress; evidence of reciprocal associations
&
\cite{nelson_resentment_2022,mcmurtrie_political_2025} \\

Affective polarization, including out-group hostility, dehumanization, moralization, and group mobilization
&
Emotional experiences and political violence risk
&
Emotional narratives, dehumanization, moralization, and group threat
&
\cite{versteegen_we_2024,piazza_political_2023} \\

Political attitudes, political conflict, election exposure, political engagement, and polarized election contexts
&
Anxiety, depression, psychological distress, sleep disturbance, and self-reported health
&
Election-related stress, political engagement, perceived policy stakes, partisan identity, election outcomes, and perceived threat
&
\cite{walker_emotional_2026,smith_pins_2024,relihan_election_2025,smith_politics_nodate,shepherd2026effects,zhao_2024_2025,pitcho-prelorentzos_factors_2018,avidor_political_2025,eshel_social_2026} \\

Ideological polarization between family members
&
Psychological need satisfaction, relationship quality, and psychological climate
&
Communication, compromise, family conflict, and relationship quality
&
\cite{gulevich_how_2026} \\

Partisan culture and polarization-related language
&
Sentiment, emotional expression, and thematic language
&
Computational linguistic proxies, political framing, and emotional expression
&
\cite{pendse_role_2025} \\

Political orientation and participation
&
Life satisfaction
&
Political interest
&
\cite{gul_tekdal_life_2026} \\

\SetCell[c=4]{c}{\textbf{Mental wellbeing $\rightarrow$ Polarization}} \\

Affective polarization, social distancing, ideological extremity, partisanship, and socio-cognitive polarization
&
Cognitive flexibility, negative interpersonal emotions, integrative complexity, and related cognitive characteristics
&
Cognitive rigidity, belief updating, integrative complexity, social rigidity
&
\cite{kam_hurt_2025,zmigrod_role_2020,salvi_does_2023,wu_framework_2022} \\

Social exclusion, ostracism, and repressive environments
&
Social exclusion and related psychological experiences
&
Group identification, perceived threat, radicalization, and support for political violence
&
\cite{pfundmair_how_2024,nugent_psychology_2017} \\

Polarized trust judgments
&
Need for cognitive closure
&
Uncertainty reduction, simplified social evaluations, and cognitive closure
&
\cite{acar-burkay_trusting_2014} \\

Political attitudes, ideological positions, and related political beliefs
&
Clinical anxiety, depression, stress, uncertainty, and perceived lack of control
&
Threat sensitivity, rumination, negativity bias, uncertainty, and perceived lack of control
&
\cite{helminen_clinical_2022,bernardi_cognitive_2024,van_prooijen_belief_2018,baum_political_2024} \\

Affective polarization, including in-group affinity and out-group hostility
&
Narcissistic personality traits
&
Rivalry-based narcissism, in-group loyalty, and inter-group hostility
&
\cite{tilley_narcissism_2025} \\

Ideological polarization estimated from online interactions
&
Cognitive distortions expressed in language
&
Computational linguistic markers of cognitive distortion and ideological extremity
&
\cite{edinger_cognitive_2025} \\

Ideological homophily and echo chamber formation
&
Preference for belief consistency
&
Belief consistency, social networks, and issue-based polarization
&
\cite{baumgaertner_preference_2022} \\

Political preferences, political orientation, voting patterns, and political or social trust
&
Depressive and anxiety-related symptoms
&
Internalizing and externalizing symptom trajectories
&
\cite{fluit_mental_2026} \\

\end{longtblr}
\twocolumn

Quantifying the association between political polarization and mental wellbeing typically involves survey-based, longitudinal, and digital-trace approaches. Large-scale surveys are the most widely used, some using standardized symptom scales \cite{helminen_clinical_2022}, and others relying on self-reported diagnoses, single-item indicators, or counts of days in poor health \cite{nayak_is_2021, fraser_harmful_2022}. Additionally, longitudinal designs have been used to track temporal ordering and individual changes in wellbeing around political events and shifts in perceived polarization \cite{relihan_election_2025, helminen_clinical_2022, lee_social_2022}. Studies also focus on cross-sectional associations between polarization, stress, and social interaction patterns \cite{nayak_is_2021, fraser_harmful_2022, panagopoulos_bridging_2022}. Furthermore, recent work incorporates mediation frameworks to test indirect pathways through which political perception affects mental wellbeing, such as reductions in social trust and weakened social connectedness \cite{reynolds_college_2025, avidor_political_2025, ma2024perceived}. For instance, Ma et al. \cite{ma2024perceived} showed that individuals’ perceptions of social fairness and confidence in government mediate the relationship between governance quality and wellbeing, highlighting the role of social trust in mental wellbeing outcomes.

Evidence directly linking specific forms of polarization to mental wellbeing is concentrated particularly around perceived and affective polarization, with more limited and heterogeneous evidence for other forms. Perceived political division
has been associated with more poor-health days and a greater likelihood of anxiety, depression and sleep disorders \cite{nayak_is_2021, fraser_harmful_2022, panagopoulos_bridging_2022}, while affective polarization has been linked to perceived stress, reduced social support and poorer self-rated health \cite{nelson_resentment_2022,mcmurtrie_political_2025}.
Political disagreement may also have interpersonal consequences: ideological differences within families can reduce constructive communication and undermine mental wellbeing, contributing to poorer relationship quality \cite{gulevich_how_2026}. More broadly, awareness of political divisions and exposure to politically consequential events have been associated with psychological distress, mood disturbances, anxiety, and disrupted sleep, and may contribute to the development of broader mental wellbeing conditions \cite{nayak_is_2021, makri_anxiety_2024, smith_pins_2024}. Qualitative research similarly reveals experiences of distress and feelings of being hurt or misunderstood, highlighting the increasing significance of political polarization as a public health concern \cite{versteegen_we_2024, kam_hurt_2025}.  At higher levels, affective and destructive forms of polarization are also linked to civic breakdowns and population stress, and, in extreme cases, dehumanization and tolerance of political violence \cite{piazza_political_2023, baum_political_2024}. 

The reverse pathway is less directly established. Emerging evidence also suggests that psychological states may shape political attitudes, with clinically assessed anxiety linked to greater concern about economic inequality, environmental issues, and political distrust \cite{helminen_clinical_2022}, while rivalry-based narcissism has been associated with stronger in-group loyalty and hostility toward political out-groups \cite{tilley_narcissism_2025}.  Related research links conspiracy beliefs with stress, uncertainty, and perceived lack of control, and depression with political attitudes through rumination and negativity bias \cite{van_prooijen_belief_2018, bernardi_cognitive_2024}. However, many of these studies do not measure political polarization directly. Instead, they examine psychological characteristics in relation to political attitudes, ideological extremity, radicalization, or other adjacent constructs. They therefore provide evidence for plausible psychological antecedents of polarization, but not equivalent evidence for a \textit{mental wellbeing $\rightarrow$ Polarization} pathway.

\begin{figure}[h!]
    \centering
    \includegraphics[width=1\linewidth]{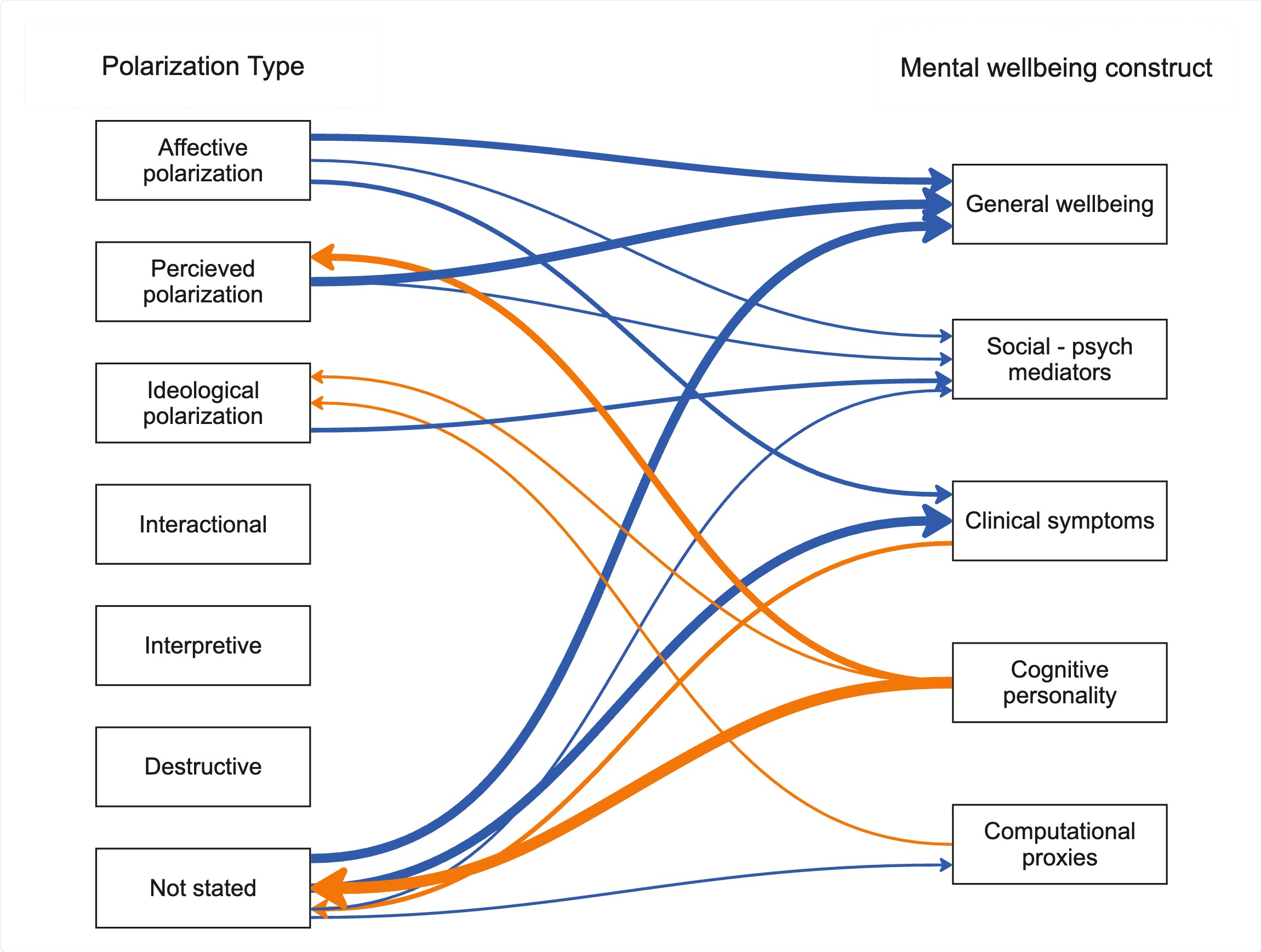}
    \caption{Conceptual overview of the bidirectional relationship between political polarization and mental wellbeing. Arrow direction indicates the direction of association, while color differentiates the two overarching pathways: blue for \textit{polarization → mental wellbeing} and orange for \textit{mental wellbeing → polarization}. Arrows represent relationships explicitly identified in the included studies. Arrow width represents the relative number of studies supporting each association.}
    \label{fig:interactions}
\end{figure}

Taken together, the evidence base is asymmetric in both volume and directness. Research on \textit{polarization $\rightarrow$ mental wellbeing} more often connects recognizable polarization measures or politically polarized contexts with psychological outcomes. In contrast, the reverse literature relies more heavily on psychological antecedents and polarization-adjacent political outcomes. This conceptual and measurement mismatch limits direct comparison between the two pathways and means that evidence consistent with bidirectionality should not be interpreted as demonstrating reciprocal effects of equivalent constructs. Figure \ref{fig:interactions} summarizes this asymmetry and the relative strength of evidence across the reviewed associations.

\subsection*{Contextual and psychological mechanisms}

\begin{figure*}[h!]
    \centering
    \includegraphics[width=0.8\linewidth]{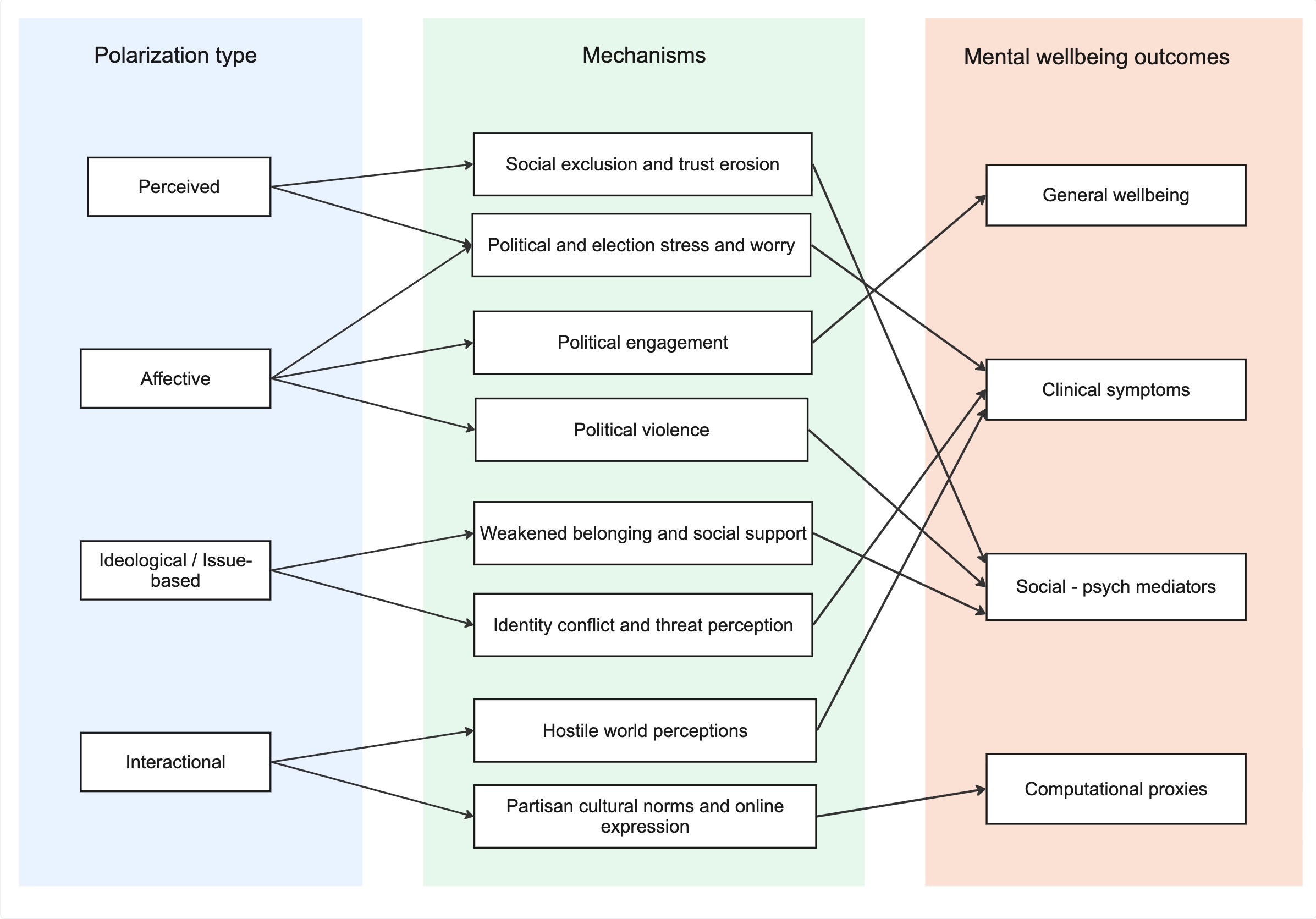}
    \caption{Schematic summary of the political polarization → mental wellbeing pathway. Arrows indicate the general direction of associations across reviewed studies. Social psychological mediators are grouped into broader social relational and affective or threat-related mechanisms. Network-based measures are treated as interactional polarization where they capture relational or interactional divisions, even when not explicitly labelled as such. See Supplementary Table 3 for specific evidence for each pathway.}
    \label{fig:polarization_wellbeing}
\end{figure*}

\begin{figure*}[h!]
    \centering
    \includegraphics[width=0.8\linewidth]{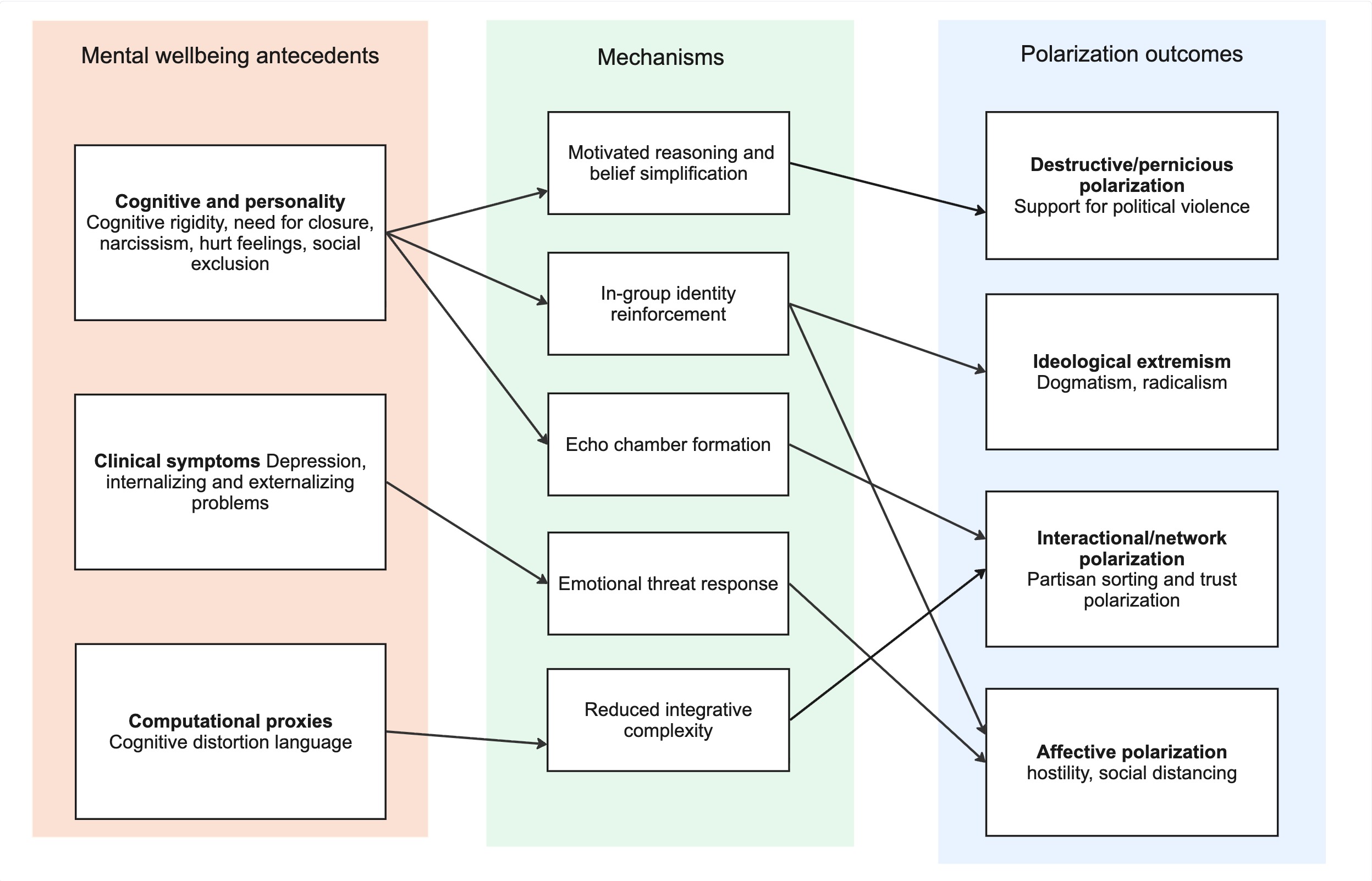}
    \caption{Schematic summary of the mental wellbeing → political polarization pathway.  Arrows indicate the general direction of associations within this pathway and reflect the collective pattern of findings across reviewed studies; the diagram is a schematic summary and does not imply that all connections are equally or independently supported by empirical evidence. See Supplementary Table 3 for specific evidence on each pathway.}
    \label{fig:wellbeing_polarization}
\end{figure*}

Identifying the mechanisms underlying these associations is essential for understanding how polarization and mental wellbeing may influence one another across different contexts. The main mechanisms identified across these studies are summarised in Figures \ref{fig:polarization_wellbeing} and \ref{fig:wellbeing_polarization}, which illustrate the pathways linking polarization and mental wellbeing in each direction.

In the \textit{Polarization $\rightarrow$ mental wellbeing} direction, the evidence suggests that the most consequential forms of polarization for mental wellbeing are those that reshape perceived social relations rather than ideological disagreement alone. Across studies, the mechanisms most frequently implicated involve disruptions to social connection, belonging, and perceived social cohesion \cite{gulevich_how_2026, smith_pins_2024, reynolds_college_2025, panagopoulos_bridging_2022}, indicating that the mental wellbeing relevance of polarization may lie primarily in its capacity to alter everyday interpersonal and community-level relationships. In this framing, polarization operates less as a political attitude structure and more as a social environment that can become chronically stressful under conditions of sustained political conflict.

Several studies identify psychological mechanisms that may shape how individuals respond to politically polarized events, particularly through perceived threat, political identity, and the personal significance attributed to political outcomes. Research on intergroup relations indicates that perceived threat and felt insecurity are associated with greater anxiety and negative affect \cite{huddy_threat_2005}. Similarly, drawing on Social Cognitive Theory, election-based studies show elevated psychological distress following politically consequential events, particularly among individuals whose preferred candidate loses \cite{pitcho-prelorentzos_factors_2018}. During the COVID-19 pandemic, partisan identity cues also shaped emotional responses and risk perceptions, demonstrating how polarization can structure divergent interpretations of shared events \cite{gadarian_partisanship_nodate, grimalda2023politicized}. Consistent with the role of identity-related threat in polarization, recent longitudinal evidence further shows that political social identity threat predicts subsequent increases in affective polarization \cite{mcmurtrie_political_2026}.

Experimental evidence further shows that anonymity and heightened identity salience increase conformity to group norms in computer-mediated settings, producing compliance under hostile norms but genuine attitude change under cooperative ones, illustrating both prosocial and antisocial forms of online polarization \cite{sabadini_compliance_2022}. Social media trace-data research shows similar processes at scale, in patterns of network segregation, selective interaction, and affective hostility \cite{tornberg_how_2022, falkenberg_patterns_2024, eck_opposing_2024, kligler-vilenchik_interpretative_2020, marchal_be_2022, phillips_high-dimensional_2023}.

Overall, findings in the literature suggest that political polarization can function as a contextual psychosocial stressor that influences mental wellbeing through sustained exposure to identity-based conflict, perceived threat, and reduced social trust \cite{finkel_political_2020, jost_cognitivemotivational_2022, kubin_role_2021}. However, these effects are also context-dependent. For example, partisan cultural environments shape how individuals experience and express distress, as well as how they seek support within social and online communities \cite{pendse_role_2025}. Conversely, contextual identity conditions can also influence the degree of polarization itself. Nugent \cite{nugent_psychology_2017} shows that repression can have different effects depending on how it is experienced: targeting specific groups can strengthen group identity and increase polarization, whereas repression experienced more broadly may weaken group identification and reduce polarization.

Beyond direct mechanisms between polarization and mental wellbeing, studies highlight shared psychological and social mechanisms that may contribute to the relationship in both directions. Perceived polarization may erode social trust, which can undermine mental wellbeing and contribute to a cycle of increasing division and alienation \cite{lee_social_2022}. Conversely, individual psychological characteristics, including cognitive inflexibility, social exclusion experiences, and personality dispositions, have been shown to foster more rigid or extreme political attitudes \cite{zmigrod_role_2020, pfundmair_how_2024, salvi_does_2023}. Cognitive processes such as belief reinforcement and motivated bias may further maintain polarized attitudes \cite{baumgaertner_preference_2022}, although their relationship with mental wellbeing remains insufficiently explored. 

In contrast, the reverse direction, whereby mental wellbeing may influence polarization, is characterized by broader cognitive and socio-emotional factors that may shape how individuals engage with political difference. However, these mechanisms are rarely examined within explicit models of political polarization and are typically studied independently, making it difficult to determine how they relate to different forms of polarization. Emerging computational research suggests that digital interactions may reinforce shared linguistic and cognitive patterns, suggesting another potential mechanism linking digital communication, political polarization, and mental wellbeing that remains largely unexplored \cite{van_papendrecht_generative_2026}. More broadly, studies across both directions use different conceptualisations, levels of analysis, and outcome measures, limiting the integration of findings into a coherent bidirectional framework. Even in related areas of political behavior, longitudinal evidence examining reciprocal associations with mental health remains limited \cite{prati_reciprocal_2024}. Consequently, although reciprocal relationships between political polarization and mental wellbeing are theoretically plausible, the current evidence remains fragmented and provides limited understanding of the mechanisms linking the two over time.

Collectively, the literature suggests a bidirectional relationship between psychological functioning and political polarization. However, bidirectionality does not necessarily imply that the mechanisms operate through the same pathways but in different directions. That is, the polarization and psychological constructs might not be aligned such that for every mechanism going from one polarization construct to a mental wellbeing construct there is a matching mechanism taking the same pathway back to the same polarization construct. Further, the individuals who are affected by the two mechanisms might be different. When there are aligned pathways, the same constructs or individuals may be involved in both directions, creating the possibility of feedback loops and cumulative effects.
Even when the pathways are not aligned, there can be reinforcing cycles because of the dynamics operating within each domain. 
Polarization may contribute to subsequent polarization through processes such as repeated exposure to political information \cite{lecheler_effects_2015}, social reinforcement through group discussion \cite{isenberg1986group}, and strengthening of political group identities \cite{renstrom2021intergroup}. Similarly, poorer mental wellbeing may contribute to persistent stress, rumination \cite{lc_rumination_nodate, hasegawa_how_2018, whisman_longitudinal_2020}, and reduced social support \cite{flynn_examination_2010}, which may in turn further impair wellbeing.

\section*{Research gaps and future directions}

Several conceptual and methodological gaps limit the integration of research on political polarization and mental wellbeing. Polarization is inconsistently defined and often not clearly distinguished into subtypes, making findings difficult to compare across studies. Measurement also varies across levels of analysis, with polarization typically assessed at the group or contextual level and mental wellbeing at the individual level through self-report or proxy measures. The predominance of cross-sectional and short-term observational designs further limits understanding of directionality and potential feedback processes. In addition, both domains are often operationalized through indirect proxies rather than validated measures, while proposed mechanisms such as social connection, threat sensitivity, and cognitive rigidity remain insufficiently tested in integrated models.

Future research should address these gaps by explicitly defining and operationalizing polarization subtypes within unified empirical designs. Multilevel approaches could link individual psychological measures to relational polarization and broader political context, while longitudinal and repeated-measures designs could clarify bidirectional and dynamic relationships. These approaches could also examine distinct and shared pathways, within-domain processes, and potential reinforcing feedback loops. Proxy measures should be validated against established polarization measures and clinical mental health instruments. Integrated mediation and moderation models could further examine how psychological processes connect polarization and mental wellbeing outcomes. Finally, expanding research beyond Western, highly educated, industrialized, and predominantly two-party contexts to multiparty and non-Western settings would improve the generalizability of findings.

\section*{Conclusion}

This review has synthesized the emerging literature on the bidirectional relationship between political polarization and mental wellbeing. Across studies, evidence indicates that political polarization is meaningfully associated with psychological functioning through cognitive, emotional, and social processes. The literature suggests that polarization is not only a political phenomenon but also a psychologically relevant social context that can influence individual wellbeing, emotional experiences, and interpersonal relations. At the same time, psychological traits and mental wellbeing-related processes may shape how individuals perceive, interpret, and respond to political differences, indicating possible reciprocal influences between the two domains. These processes involve cognitive, emotional, and social mechanisms that operate across individual and contextual levels. Multiple levels of influence, including personality, identity, and environmental factors, therefore need to be considered when examining the relationship between polarization and mental wellbeing.

Taken together, the evidence remains fragmented across disciplines and is largely correlational, with inconsistent conceptualizations, misaligned measurement approaches, and limited longitudinal and mechanistic evidence. These gaps indicate that the field remains in an early integrative stage. Advancing it will require greater conceptual standardization, better alignment of polarization and psychological measures, longitudinal and multilevel designs, and explicit testing of psychological mechanisms within unified bidirectional frameworks. Such approaches are needed to clarify when, how, and under what social and political conditions polarization and mental wellbeing influence one another.

\bibliographystyle{IEEEtran}
\bibliography{references}

\section*{Funding}
This work was supported by Research Council of Finland through projects POLEMIC (371535 and 371536), and PolWell (368910). S.S acknowledges support by the Helsinki Institute of Information Technology (HIIT).

\section*{Author Contributions}
A.M.: conceptualization, methodology, formal analysis, data curation, writing—original draft, writing—review and editing, visualization.
S.S.: conceptualization, writing—review and editing.
J.K.: conceptualization, writing—review and editing, supervision.
M.K.: conceptualization, writing—review and editing,
supervision.
T.A.: conceptualization, writing—review and editing,
supervision, project administration.

\section*{Competing interests}
The authors declare no competing interests.

\section*{Supplementary Tables}

\setcounter{table}{0}
\renewcommand{\tablename}{Supplementary Table}
\renewcommand{\thetable}{\arabic{table}}
\onecolumn
\singlespacing
\begin{longtblr}[
caption = {Conceptual Frameworks and Dimensions of Political Polarization},
label = {tblr:something},
]{
colspec = {Q[l] Q[l] Q[l] Q[l] Q[l] Q[l]},
colspec = {XXXXXX}, width = \textwidth,
rowhead = 1,
hlines={black},
row{1} = 
}
\textbf{Author \& Year} & \textbf{Focus} & \textbf{Concept of Polarization} & \textbf{Dimension} & \textbf{Population Studied} & \textbf{Main Findings} \\

McCoy et al. (2018) \cite{mccoy_polarization_2018} & Explain democratic dangers of polarization & Polarization as an identity-based “us vs. them” divide & Affective and ideological polarization & Cross-national democratic systems & Polarization erodes democratic norms and fosters identity-driven partisanship \\

Iyengar et al. (2019) \cite{iyengar_origins_2019} & Identify psychological bases of polarization & Affective polarization as growing dislike and distrust between partisans & Affective and perceptual mechanisms & U.S. voters across multiple national surveys & Affective polarization rooted in identity alignment and negative partisanship \\

Druckman \& Levendusky (2019) \cite{druckman_what_2019} & Review affective polarization and how to measure it & Affective polarization leads to stronger in-group preference and bias & Affective polarization & U.S. voters & Most measures align, but social-distance differs, and party attitudes mainly reflect dislike of elites \\

Kubin \& von Sikorski (2021) \cite{kubin_role_2021} & Examine the media’s contribution to polarization & Polarization as ideological divergence and emotional hostility & Media exposure, affective and ideological polarization & Western democracies and media audiences & Media fragmentation intensifies perceived polarization and partisan animosity, but it also has the potential to foster depolarization \\

Wilson (2020) \cite{wilson_polarization_2020} & Examine societal-level contributors to political polarization & Differentiates ideological, affective, and perceived polarization & Political elites, partisan media, social media dynamics & Theoretical synthesis (no primary data) & Institutional factors create perceived divisions, driving affective and ideological polarization. \\

Yang et al. (2016) \cite{yang_why_2016} & Explore polarization in digital media environments & Perceived polarization occurs when individuals see political or social groups as distant in their policy preferences & Perceived polarization and media & Adult citizens across ten countries using representative national surveys & Online news consistently relates to perceived polarization, while traditional media effects vary by country \\

Campos (2025) \cite{campos_new_2025} & Measuring affective polarization & Affective polarization as the focus on emotional and social distances between partisans & Affective polarization, othering, aversion, and moralization & Theoretical synthesis (no primary data) & Developed a nine-item measure capturing distinct components of affective polarization \\

McMurtrie (2024) \cite{mcmurtrie_development_2024} & Development of a scale to assess affective polarization & Affective polarization as expressed dislike and negative feelings toward political out-groups & Affective polarization, social distance, aversion, and incivility & U.S. adults & Developed a three-factor scale capturing distinct aspects of affective polarization. \\

Frimer (2023) \cite{frimer_incivility_2023} & Analyze incivility in Twitter & Polarization expressed through uncivil political discourse & Incivility in political communication, linguistic indicators & U.S. politicians on Twitter & Incivility rose 23\% over a decade, fueled by positive feedback on likes and retweets. \\

Fernbach (2022) \cite{fernbach_false_2022} & Review empirical research on perceived polarization and negative meta-perceptions & Polarization includes policy, affective, social, and factual divides & Perceived polarization & Theoretical
synthesis(no primary data) & Perceived polarization stems from categorical thinking, projection, and affective reasoning.\\

Moore-Berg (2020) \cite{moore-berg_exaggerated_2020} & Examine how exaggerated meta-perceptions drive partisan hostility & Polarization as emotional distancing and hostility between political groups & Perceived and affective polarization & U.S. political partisans & Exaggerating opponents’ extremity and hostility fuels inter-group hostility and affective polarization.

\end{longtblr}
\onehalfspacing

\clearpage
\onecolumn
\singlespacing
\begin{longtblr}[
caption = {Overview of studies on polarization},
label = {tblr:something},
]{
colspec = {Q[l] Q[l] Q[l] Q[l] Q[l] Q[l]},
colspec = {XXXXXX}, width = \textwidth,
rowhead = 1,
hlines={black},
row{1} = 
}
\textbf{Study} & \textbf{Focus} & \textbf{Polarization Type} & \textbf{Methods} & \textbf{Population Studied} & \textbf{Main Findings} \\

Lelkes (2016) \cite{lelkes_mass_2016} & Review mass-level polarization, and how to measure it & Ideological consistency, ideological divergence, perceived polarization, and affective polarization & 7-point party identity scale, ideological self-placement measure, feeling thermometers and ideological distances & U.S. adults from the American National Election Studies (ANES) & Finds increasing affective but not necessarily ideological polarization \\

Törnberg (2022) \cite{tornberg_how_2022} & Model network polarization dynamics in online environments & Affective polarization & Modeling: agent-based modeling + theory from opinion dynamics, digital media interaction, and literatures & Synthetic social networks and online data & Digital media contributes to the rise of affective polarization by enabling partisan sorting and alignment of social identities \\

Falkenberg et al. (2024) \cite{falkenberg_patterns_2024} & Examine online political interaction networks and how toxic discourse reflects polarization & Structural and affective polarization & Network mapping, toxicity classification (out-group vs in-group), engagement metrics & Twitter users (politically engaged accounts) in nine countries & Out-group interactions are more toxic and less engaging, consistent global pattern of online polarization \\

Eck et al. (2024) \cite{eck_opposing_2024} & Review how climate skepticism and activism discourse evolved online & Positional, interactional, and affective polarization & Topic modeling, framing, content, sentiment and network analyses & General online media and specific online platforms & Frames online climate change polarization as six-dimensional \\

Yarchi et al. (2021) \cite{yarchi_political_2021} & Political polarization across social media platforms & Interactional, Positional, and Affective Polarization & Combines computational and manual text analysis, topic modeling, sentiment, and network homophily measures & Social media (Facebook, Twitter, and WhatsApp) users discussing political controversies in Israel & Polarization varies by platform: high homophily on Twitter, WhatsApp shows some depolarization, lowest homophily on Facebook \\

Phillips et al. (2023) \cite{phillips_high-dimensional_2023} & Propose high-dimensional models for polarization detection & Social, knowledge, and knowledge source polarization & Hybrid network and content analysis using a high-dimensional framework & Twitter users discussing COVID-19 vaccines &  Enables multidimensional analysis of online polarization to inform interventions \\

Muñoz et al. (2024) \cite{munoz_quantifying_2024} & Quantify online polarization across platforms & Not explicitly categorized; measured via information flow and community structures & Hybrid network and information algorithm (SPIN), combining temporal, content, and structural aspects & Official political party accounts on X (Twitter) & SPIN shows rising polarization over election cycles, higher in national than local elections\\

Salloum (2022) \cite{salloum_separating_2022} & Evaluate and normalize structural polarization measures in social networks & Structural polarization & Network construction, graph partitioning (e.g., METIS, spectral clustering), and EI-index & Synthetic and online social networks & Normalization enhances polarization measure performance; method choice is less critical \\

Interian et al. (2023) \cite{interian_network_2023} & Review of online network polarization measures and reduction methods & Structural polarization & Measures based on homophily, modularity, random walks, and balance theory (controversy, disagreement, conflict) & Theoretical synthesis(no primary data) & Reducing polarization through network edits (node or edge modifications) and system redesigns \\

Karjus et al. (2024) \cite{karjus_evolving_2024} & Track language change in polarized contexts & No explicitly categorized (interactional polarization) & Hybrid data-mining, semantic, and machine-learning approach with human annotation & U.S. X (Twitter) users classified by political leaning via followed media accounts & Political groups increasingly diverge in language and topics, heightening miscommunication and group separation \\

Ross Arguedas et al. (2022) \cite{ross_arguedas_echo_2022} & Review echo chamber formation and exposure diversity & Ideological, affective and interactional polarization &  Surveys, content and network analyses & Theoretical synthesis (no primary data) & Echo chambers occur less frequently than commonly believed \\

Kligler-Vilenchik et al. (2020) \cite{kligler-vilenchik_interpretative_2020} & Investigate how political polarization evolves across different social media platforms & Interpretative and positional polarization & Hybrid content analysis using structural topic modeling and manual validation & Users engaged in political discussions on Facebook, X (Twitter), and WhatsApp & Political disagreement appears early, stays stable, and varies across platforms \\

Marchal \& Au (2022) \cite{marchal_be_2022} & Investigate affective polarization in online political discussions & Affective polarization &  Hybrid sentiment analysis, ideological approach, and modeling approach to interaction patterns & Reddit users participating in r/politics discussions & Cross-ideological interactions were shorter and more negative, but positive out-group sentiment improved dialogue\\
\end{longtblr}
\onehalfspacing


\onecolumn
\begin{longtblr}[
    caption={Studies examining political polarization and mental wellbeing, including proxy measures, by direction of association, measurement approach, and key findings},
    label={tab:polarization_wellbeing_appendix}]{
    width=\linewidth,
    colspec = {Q[l] Q[l] Q[l] Q[l] Q[l] Q[l]},
    colspec={X[0.9]X[1.5]X[1]X[1.5]X[1.8]X[1.9]},
    rowhead=1,
    hlines
}

\textbf{Study} & \textbf{Method} & \textbf{Population Studied} & \textbf{Polarization Measure (Type)} & \textbf{Psychological Measures (Construct Category)} & \textbf{Main Findings} \\

\SetCell[c=6]{c}{\textbf{Polarization $\rightarrow$ mental wellbeing}} \\

Fraser (2022) \cite{fraser_harmful_2022} & National survey with quota sampling; statistical models including negative binomial and zero-inflated models & U.S. adults & Custom 0–10 perceived political distance scale; perceived political distance from average voter \newline\textbf{(Perceived polarization)}& Self-reported days of poor physical/mental health \newline\textbf{(General
psychological wellbeing)} & Greater perceived political distance from one’s state’s median voter corresponds with more days of mental health declines in those with fair mental health \\

Nayak et al. (2021) \cite{nayak_is_2021}  & Multivariable logistic regression & U.S. adults & 5-point Likert scale item measuring perceived partisan polarization change \newline\textbf{(Perceived polarization)} & Self-reported onset of anxiety, depression, and sleep disorders \newline\textbf{(General psychological wellbeing)} & Perceived and state-level polarization were linked to higher odds of depression, anxiety, and sleep disorders \\

Makri (2024) \cite{makri_anxiety_2024}  & Commentary-style review; no original empirical data & Adult respondents in political contexts & Contextual political exposure indicators \newline\textbf{(Perceived polarization)} & Anxiety, depression, headaches  \newline\textbf{(General psychological wellbeing)} & Documents relations between political stressors and elevated anxiety over time (contextual evidence) \\

Smith et al. (2024) \cite{smith_pins_2024} &  Two-wave panel survey (pre- and post-election) & U.S. adults & self reported political engagement and perceived exposure to partisan conflict \newline\textbf{(Perceived polarization)} & 8-item Political Anxiety Scale  \newline\textbf{(General psychological wellbeing)} & Anxiety peaked before the election and declined afterward overall; politically engaged individuals had higher anxiety levels \\

Versteegen (2024) \cite{versteegen_we_2024} & Semi-structured interviews; qualitative analysis &  Radical-right voters in Germany & Narrative accounts of political attitudes \newline\textbf{(Affective polarization)} & Narrative accounts of emotions \newline\textbf{(General psychological wellbeing)}  & Partisans use emotion narratives to praise the in-group, vilify the out-group, and amplify emotional divides \\

Lee (2022) \cite{lee_social_2022} & Panel and longitudinal survey analysis; fixed-effects panel regression & U.S.adults & Perceived polarization survey items \newline\textbf{(Perceived polarization)} &  Three-item generalized social trust ratings + trust in co-nationals \newline\textbf{(Social-psychological mediators)}  & Perceived political polarization lowers social trust and reduces willingness to cooperate with others \\

Piazza (2023) \cite{piazza_political_2023} & Original survey and cross-national time series analyses, logistic and time series regression models & U.S. residents and citizens from 83 democratic countries & Dehumanization, moralistic and zero-sum nature, and group mobilization \newline\textbf{(Affective polarization)} & Political violence risk; \newline\textbf{(Social-psychological mediators)} & Higher affective polarization boosts support for political violence domestically and internationally \\

Relihan (2025) \cite{relihan_election_2025} & Longitudinal panel around election events using logistic regression analyses & U.S. adults surrounding election periods & Political identity (Republican vs. Democrat) \newline\textbf{(Not explicitly stated)} & Heightened election-related worry leading to physical health outcomes \newline\textbf{(General psychological wellbeing)} & Finds temporal increases in health complaints associated with election worry; stronger effect for Republicans \\

Panagopoulos (2022) \cite{panagopoulos_bridging_2022} & National surveys; regression analyses & U.S.adults & 10 point scale at state and national levels; bonding, bridging and linking social capital \newline\textbf{(Perceived polarization)} & Number of poor physical and mental health days \newline\textbf{(General psychological wellbeing)} & Bonding ties were associated with better health in polarized settings; findings for bridging ties were mixed \\

Reynolds (2025) \cite{reynolds_college_2025} & Online survey; Ordinary least squares regression and KHB mediation models & U.S. college student sample & Subjective polarization \newline\textbf{(Not explicitly stated)}  & Social connectedness and belonging, psychological distress scales \newline\textbf{(Social-psychological mediators)} & Finds social connectedness and belonging mediate part of the association between perceived polarization and distress \\

McMurtrie et al. (2025) \cite{mcmurtrie_political_2025} & Six wave longitudinal survey; Random Intercept Cross Lagged Panel Models (RI CLPM) & U.S. adults &  Affective Polarization Scale (social distance, aversion, incivility toward political out-groups) \newline\textbf{(Affective polarization)} & Social support, perceived stress (PSS-10), self rated health \newline\textbf{(Clinical symptoms and Social-psychological mediators)} & Higher affective polarization was linked to lower social support, higher stress, and poorer health, while perceived stress predicted later increases in affective polarization \\


Pendse et al. (2025) \cite{pendse_role_2025} & NLP and statistical matching & U.S. adults, online partisan and mental health subreddit members &  Partisan culture; language indicative of political polarization \newline\textbf{(Not explicitly stated)}  & Sentiment, emotional expression, and thematic content of online posts \newline \textbf{(Computational proxies)} & Democrats tend to use more clinical and politically framed language while Republicans emphasize social and relational contexts \\

Avidor and Shmotkin (2025) \cite{avidor_political_2025}  & Survey study & Israeli adults & Political orientation; political attitude during a socio-political crisis \newline\textbf{(Not explicitly stated)} & Anxiety (GAD-7), depressive symptoms (CES-D), wellbeing (life satisfaction, positive/negative affect) \newline\textbf{(Clinical symptoms)} & No differences in overall wellbeing, but leftists showed higher distress; hostile world perceptions partially explained associations between political orientation and mental health indicators \\

Nelson (2022) \cite{nelson_resentment_2022} & Nationally representative survey (Pew American Trends Panel); structural equation modelling & U.S. adults & Negative emotions toward the opposing party (anger, fear, frustration)\newline\textbf{(Affective polarization)} & Self-rated health \newline\textbf{(General psychological wellbeing)} & Affective polarization functioned as a sociopolitical stressor and was associated with poorer self-rated health, despite a smaller positive effect through increased political participation \\

Zhao et al. (2025) \cite{zhao_2024_2025} & Two wave longitudinal survey; logistic regression & U.S. young adults & Election related stress (anticipatory, news related, election outcome) in a highly polarized election context \newline\textbf{(Not explicitly stated)} & PHQ 8 depression, GAD 7 anxiety \newline\textbf{(Clinical symptoms)} & News related stress was associated with higher risks of depression and anxiety, while anticipatory election stress was associated with higher risk of depression \\

Pitcho-Prelorentzos et al. (2018) \cite{pitcho-prelorentzos_factors_2018} & National post-election survey; regression analyses & U.S. adults & Post-election context of a highly polarized election; dimensions of political self-efficacy \newline\textbf{(Not explicitly stated)}  & Psychological distress (PHQ-4), negative affect (sadness, anxiety, anger) \newline\textbf{(Clinical symptoms)} & Through a social cognitive theory lens, election related psychological distress was higher among younger individuals, those more politically engaged, heavier users of new media, and those whose preferred candidate lost the election \\

Smith (2022) \cite{smith_politics_nodate} & Two-wave panel survey (pre- and post-election) & U.S. adults & Contextual political engagement \newline\textbf{(Not explicitly stated)} & Self-reported mental health and public health indicators \newline\textbf{(General psychological wellbeing)} & Higher political engagement during a highly polarized political period is associated with worse public health and increased psychological distress (including stress, loss of sleep, or suicidal thoughts) \\

Tekdal \& Avcil (2026) \cite{gul_tekdal_life_2026} & Cross-sectional survey (World Values Survey; moderated mediation analysis) & Türkiye adults & Political orientation, interest, and participation \newline\textbf{(Not explicitly stated)} & Self-reported life satisfaction \newline\textbf{(General psychological wellbeing)} & Political orientation moderated the indirect association between life satisfaction and political participation through political interest \\

Shepherd \& Albertson (2026) \cite{shepherd2026effects} & Observational study using Behavioral Risk Factor Surveillance System survey data and Google Trends & U.S. adults (2020 and 2024 presidential elections) & Exposure to polarized elections and partisan election outcomes \newline\textbf{(Not explicitly stated)} & Self reported poor mental health; politics related mental health search interest \newline\textbf{(General psychological wellbeing)} & Recent polarized elections were associated with poorer mental health, particularly among likely partisan election losers and individuals with greater perceived policy stakes \\

Eshel et al. (2026) \cite{eshel_social_2026} & Cross-sectional online survey & Israeli Jewish adults & Social polarization measured through ethnic identity, political orientation (government support/opposition), and religiosity (affective/social identity polarization) \newline\textbf{(Not explicitly stated)} & Psychological distress: depression symptoms, anxiety symptoms, sense of danger, and perceived threats \newline\textbf{(Clinical symptoms)} & Positions within major social divisions were associated with distress responses following collective trauma. Political orientation predicted differences in anxiety, depression, sense of danger, and perceived threat \\

Gulevich \& Semiletova (2026) \cite{gulevich_how_2026} & Cross-sectional online survey; mediation analysis & Russian adults & Ideological polarization between family members based on perceived similarity of political views \newline\textbf{(Ideological/issue-based polarization)} & Family relationship quality, psychological climate, and basic psychological need satisfaction \newline\textbf{(Social-psychological mediators)} & Ideological polarization between family members was associated with less compromise and more tense communication, which reduced psychological need satisfaction and negatively affected relationship quality \\

Walker et al. (2026) \cite{walker_emotional_2026} & Two longitudinal experience sampling (diary) studies & U.S. adults &  Political attitudes, issue-based ideological extremity, feeling thermometer difference \newline\textbf{(Ideological/issue-based polarization and Affective polarization)} & GAD-7 anxiety symptoms \newline\textbf{(Clinical symptoms)} & Greater political engagement was linked to increased affective instability, which in turn was associated with higher anxiety. These effects were stronger among individuals with higher ideological extremity and affective polarization \\
 
\SetCell[c=6]{c}{\textbf{mental wellbeing $\rightarrow$ Polarization}} \\


Kam (2025) \cite{kam_hurt_2025} & Cross-sectional survey study with correlational and regression analyses & U.S. adults & Feeling thermometer, Bogardus Social Distance scale, political identification \newline\textbf{(Affective polarization)} & 5 negative interpersonal feelings, integrative complexity (AutoIC) \newline\textbf{(Cognitive \& personality factors)} & Emotional wounds correlate with greater affective polarization and social distancing, and lower complexity in political reasoning. \\

Wu et al. (2022) \cite{wu_framework_2022} & Conceptual synthesis integrating psychological and sociopolitical literature & Theoretical (no sampled population) & Conflict mindset, group conformity, and simplification of information \newline\textbf{(Not explicitly stated)} & Cognitive inflexibility, belief updating, integrative complexity, rigid cognitive specialization \newline\textbf{(Cognitive \& personality factors)} & Polarization fosters cognitive rigidity, reinforcing group conformity and belief simplification through a self-reinforcing feedback loop  \\

Zmigrod (2020) \cite{zmigrod_role_2020} & Conceptual review & Theoretical (no sampled population) & Ideological extremism, partisanship, and dogmatism \newline\textbf{(Not explicitly stated)} & Cognitive inflexibility \newline\textbf{(Cognitive \& personality factors)} & Cognitive rigidity associated with ideological extremism; bidirectional links between ideology and cognition are proposed \\

Pfundmair et al. (2024) \cite{pfundmair_how_2024} & Conceptual review & Theoretical (no sampled population) & Social exclusion/ostracism measures \newline\textbf{(Not explicitly stated)} & Radicalism intentions/approval of extreme actions and violence \newline\textbf{(Cognitive \& personality factors)}  & Social exclusion increases willingness to support extreme political violence and radical ideologies \\

Acar-Burkay et al. (2014) \cite{acar-burkay_trusting_2014} & Six experiments manipulating situational and chronic need for cognitive closure across varying interaction contexts & Graduate students & Interpersonal trust (generalized trust judgments) \newline\textbf{(Not explicitly stated)}  & Need for closure scale \newline\textbf{(Cognitive \& personality factors)}  & Higher need for closure is associated with more polarized trust judgments, as individuals reduce uncertainty by relying on simplified evaluations of others\\

Edinger et al. (2025) \cite{edinger_cognitive_2025} & Longitudinal social media text analysis of Twitter posts & Twitter users discussing the 2016 and 2020 elections & Latent ideological estimation using retweet interactions \newline\textbf{(Ideological/issue-based polarization)}  & Lexical markers of the expression of cognitive distortions in language \newline\textbf{(Computational
proxies)} & Cognitive distortion language increased over time, with ideological differences in baseline levels and in its association with extremism, suggesting a potential link to political polarization\\ 

Nugent (2017) \cite{nugent_psychology_2017} & Lab experiment plus comparative case analysis & Tunisian adults in the experiment; opposition groups in Egypt and Tunisia in case studies & In-group identification, out-group attitudes, preference polarization \newline\textbf{(Affective Polarization)} & Primes of repressive environments  \newline\textbf{(Cognitive \& personality factors)} & Targeted repression increases polarization by reinforcing in-group identity, while widespread repression reduces polarization and strengthens out-group identification\\

Tilley et al. (2025) \cite{tilley_narcissism_2025} & Nationally representative two-wave panel survey & British adults (party and Brexit identities) & In-group affinity and out group animosity measured through partisanship and stereotype \newline\textbf{(Affective Polarization)} & 18-item Narcissistic Admiration and Rivalry Questionnaire (NARQ) (admiration vs. rivalry) \newline\textbf{(Cognitive \& personality factors)} & Higher narcissism  predicts both in-group loyalty and hostility toward political out-groups \\

Salvi et al. (2023) \cite{salvi_does_2023} & Survey with cognitive tasks & U.S. adults & Four latent profiles of social rigidity (Socio-cognitive polarization, bullshit receptivity, overclaiming, cognitive rigidity) \newline\textbf{(Not explicitly stated)} & Cognitive flexibility through problem solving \newline\textbf{(Cognitive \& personality factors)} & Individuals with lower social rigidity performed the best on problem-solving, showing that social and cognitive rigidity reinforce each other \\

Baum et al. (2024) \cite{baum_political_2024} & Two-wave national survey panel & U.S. adults &  Support for election violence/riots; conspiracy beliefs/participatory inclinations, feeling thermometer \newline\textbf{(Not explicitly stated)} & Depressive symptoms, PHQ-9\newline\textbf{(Clinical symptoms)} & Depression was associated with greater support for political violence among individuals with conspiracy beliefs and/or strong participatory inclinations \\

Baumgaertner et al. (2022) \cite{baumgaertner_preference_2022} & Computational simulation and network modelling informed by survey data on ideological positions and online social networks  & U.S. adults & ideological homophily and echo chamber formation \newline\textbf{(Ideological/issue-based polarization)} & Preference for belief consistency \newline\textbf{(Cognitive \& personality factors)} & Even without traditional exclusion mechanisms, echo chambers and issue polarization can emerge through the interaction between belief preferences and social networks \\

Fluit et al. (2026) \cite{fluit_mental_2026} & Three-decade longitudinal cohort study & Norwegian individuals followed from adolescence to midlife & Political preferences, political orientation, voting patterns, political/social trust \newline\textbf{(Not explicitly stated)} & adolescent internalizing (depression/anxiety) and externalizing problems \newline\textbf{(Clinical symptoms)} & Higher internalizing and externalizing symptom trajectories were associated with lower political and social trust and distinct ideological profiles in midlife  \\
\end{longtblr}

\end{document}